\documentclass[review]{elsarticle}

\usepackage{lineno,hyperref}

\journal{arXiv}
\usepackage{amsmath,amssymb,amsfonts}
\usepackage{algorithmic}
\usepackage{graphicx}
\usepackage{textcomp}
\usepackage{xcolor}
\usepackage{caption}
\usepackage{subcaption}
\usepackage{booktabs}
\usepackage[T1]{fontenc}
\usepackage{amsmath}
\usepackage{balance}
\usepackage{xurl,multirow}

\begin{document}

\begin{frontmatter}

\title{Towards an Effective and Efficient Deep Learning Model for COVID-19 Patterns Detection in X-ray Images}


\author[decomufop]{Eduardo Luz}
\author[ppgccufop]{Pedro Silva}
\author[decomufop]{Rodrigo Silva}
\author[hos]{Ludmila Silva}
\author[decomufop]{Gladston Moreira \corref{mycorrespondingauthor}}
\ead[url]{http://www.decom.ufop.br/csilab/}
\cortext[mycorrespondingauthor]{Corresponding author}
\ead{gladston@ufop.edu.br}
\author[parana]{David Menotti}

\address[decomufop]{Computing Department, Federal University of Ouro Preto, Campus Morro do Cruzeiro, Ouro Preto-MG, Brazil}
\address[ppgccufop]{Graduate Program in Computer Science, Federal University of Ouro Preto (UFOP), MG, Brazil}
\address[hos]{Hospital e Maternidade Otaviano Neves, Belo Horizonte-MG, Brazil}
\address[parana]{Department of Informatics, Federal University of Paran\'a, Centro Polit\'ecnico, Jardim das Am\'ericas, Curitiba-PR, Brazil}

\begin{abstract}
{\bf Purpose:} Confronting the pandemic of COVID-19, is nowadays one of the most prominent challenges of the human species. A key factor in slowing down the virus propagation is the rapid diagnosis and isolation of infected patients. The standard method for COVID-19 identification, the Reverse transcription polymerase chain reaction method, is time-consuming and in short supply due to the pandemic. Thus, researchers have been looking for alternative screening methods and deep learning applied to chest X-rays of patients has been showing promising results. Despite their success, the computational cost of these methods remains high, which imposes difficulties to their accessibility and availability. Thus, the main goal of this work is to propose an accurate yet efficient method in terms of memory and processing time for the problem of COVID-19 screening in chest X-rays.\\
{\bf Methods:} To achieve the defined objective we exploit and extend the EfficientNet family of deep artificial neural networks which are known for their high accuracy and low footprints in other applications. We also exploit the underlying taxonomy of the problem with a hierarchical classifier. A dataset of 13,569 X-ray images divided into healthy, non-COVID-19 pneumonia, and COVID-19 patients is used to train the proposed approaches and other 5 competing architectures. Finally, 231 images of the three classes were used to assess the quality of the methods.\\
{\bf Results:} The results show that the proposed approach was able to produce a high-quality model, with an overall accuracy of  93.9\%, COVID-19, sensitivity of 96.8\% and positive prediction of 100\%, while having from 5 to 30 times fewer parameters than other than the other tested architectures. Larger and more heterogeneous databases are still needed for validation before claiming that deep learning can assist physicians in the task of detecting COVID-19 in X-ray images.\\
{\bf Conclusions:} We believe the reported figures represent state-of-the-art results, both in terms of efficiency and effectiveness, for the COVID$x$ database, a database comprised of 13,800 X-ray images, 183 of which are from patients affected by COVID-19. The current proposal is a promising candidate for embedding in medical equipment or even physicians' mobile phones.
\end{abstract}

\begin{keyword}
COVID-19\sep Deep Learning\sep EfficientNet\sep Pneumonia\sep Chest (X-ray) Radiography.
\end{keyword}

\end{frontmatter}


\section{Introduction}
\label{sec:introduction}


The COVID-19 is an infection caused by the SARS-CoV-2 virus and may manifest itself as a flu-like illness potentially progressing to an acute respiratory distress syndrome. The disease severity resulted in a global public health efforts to contain person-to-person viral spread by early detection \cite{davarpanah2020novel}.

The Reverse-Transcriptase Polymerase Chain Reaction (RT-PCR) is currently the gold standard for a definitive diagnosis of COVID-19. However, false negatives have been reported (due to insufficient cellular content in the sample or inadequate detection and extraction techniques) in the presence of positive radiological findings \cite{araujo2020covid}. Therefore, effective exclusion of the COVID-19 infection requires multiple negative tests, possibly exacerbating test kit shortage \cite{american2020acr}. 

As COVID-19 spreads in the world, there is growing interest in the role and suitability of chest X-Rays (CXR) for screening, diagnosis, and management of patients with suspected or known COVID-19 infection \cite{huang2020use, ai2020correlation, ng2020imaging}. Besides, there have been a growing number of publications describing the CXR appearance in patients with COVID-19 \cite{american2020acr}.

The accuracy of CXR diagnosis of COVID-19 infection strongly relies on radiological expertise due to the complex morphological patterns of lung involvement which can change in extent and appearance over time. The limited number of sub-specialty trained thoracic radiologists hampers reliable interpretation of complex chest examinations, specially in developing countries, where general radiologists and occasionally clinicians interpret chest imaging \cite{davarpanah2020novel}.

Deep Learning is a subset of machine learning in artificial intelligence (AI) concerned with algorithms inspired by the structure and function of the brain called artificial neural networks. 
Since deep learning techniques, in particular convolutional neural networks (CNNs), have been beating humans in various computer vision tasks~\cite{lecun2015deep, touvron2020fixing, rajpurkar2017chexnet}, it becomes a natural candidate for the analysis of chest radiography images. 

Deep learning has already been explored for the detection and classification of pneumonia and other diseases on radiography \cite{rajpurkar2017chexnet,wang2017hospital,jaiswal2019identifying}. 
In this context, this work aims to investigate deep learning models that are capable of finding patterns X-ray images of the chest, even if the patterns are imperceptible to the human eye, and to advance on a fundamental issue: computational cost. 

Finding a model of low computational cost is important because it allows the exploitation of input images of much higher resolutions without making the processing time prohibitive. Besides, it becomes easier and cheaper to embed these models in equipment with more restrictive settings such as smartphones. We believe that a mobile application, that integrates deep learning models for the task of recognizing patterns in x-rays must be easily accessible and readily available to the medical staff. For such aim, the models must have a low footprint and low latency, that is, the models must require little memory and perform inference quickly to allow use on embedded devices and large scale, enabling integration with smartphones and medical equipment. 

To find such cost-efficient models, in this work, the family of EfficientNets, recently proposed in \cite{tan2019efficientnet}, is investigated. These models have shown high performance in the classic ImageNet dataset \cite{imagenet_cvpr09} while presenting only a small fraction of the cost of other popular architectures such as the ResNets and VGGs. We also exploit the natural taxonomy of the problem and investigate the use of hierarchical classification. In this case, low computational cost is even more critical since multiple models have to be built.


The results show that it is indeed possible to build much smaller models without compromising accuracy. Thus, embedding the proposed neural network model in a mobile device to make fast inferences becomes more feasible. Despite its low computational cost, the proposed model achieves high accuracy (93.9\%) and detect infection caused by COVID-19 on chest X-rays with a Sensitivity of 96.8\% and Positivity Prediction of 100\% (without false positives). Regarding the hierarchical model, in addition to consuming more computational resources than the flat classification, it showed to be less effective for minority classes, which is the case for the COVID-19 class in this work. However, we believe that the hierarchical method is the most suitable for the application. It may suffer less from bias in the evaluation protocols~\cite{maguolo2020critic}, since images from different sources (datasets) are mixed to build the superclasses according to the taxonomy of the problem
~\cite{silla2011survey}.

The development of this work may allow the future construction of an application for use by the medical team, through a camera on a regular cell phone. The source code as the pre-trained models are available in~\url{https://github.com/ufopcsilab/EfficientNet-C19}. 

The remainder of this work consists of six sections. 
Section~\ref{sec:related} presents a review of related works.
Section~\ref{sec:setting} defines the problem tackled in this paper. 
The methodology and the dataset are described in Section~\ref{sec:approach}.
In Section~\ref{sec:experiments}, the results of a comprehensive set of computational experiments are presented. 
In Section~\ref{sec:future}, propositions for future research in the area are addressed. Finally, conclusions are pointed out in Section~\ref{sec:conclusion}.

\section{Related Works}
\label{sec:related}

Addressing the COVID-19, in \cite{hemdan2020covidx}, a comparison among seven different well-known deep learning neural networks architectures was presented. 
In the experiments, they use a small dataset with only 50 images in which 25 samples are from healthy patients and 25 from COVID-19 positive patients. 
The models were pre-trained with the ImageNet dataset \cite{imagenet_cvpr09}, which is a generic image dataset with over 14 million images of all sorts, and only the classifier is trained with the radiography. 
In their experiments, the VGG19 \cite{simonyan2014very} and the DenseNET201 \cite{huang2019gpipe} were the best performing architectures. 

In \cite{wang2020covidnet}, a new architecture of CNN, called COVID-net, is created to classify CXR images into normal, pneumonia, and COVID-19. Differently from the previous work, they use a much larger dataset consisting of $13,800$ CXR images across $13,645$ patient cases from which $182$ images belong to COVID-19 pacients.
The authors report an accuracy of 92.4\% overall and sensitivity of 80\% for COVID-19. 

In \cite{farooq2020covid}, the ResNet50 \cite{szegedy2017inception} is fine-tuned for the problem of classifying CXRs into normal, COVID-19, bacterial-pneumonia and viral pneumonia. The authors report better results when compared with the COVID-net, 96.23\% accuracy overall, and 100\% sensitivity for COVID-19. Nevertheless, it is important to highlight that the problem in \cite{farooq2020covid} has an extra class and that its dataset is a subset of the dataset used in \cite{wang2020covidnet}. In \cite{farooq2020covid} the dataset consists of 68 COVID-19 radiographs from 45 COVID-19 patients, 1,203 healthy patients, 931 patients with a bacterial pneumonia and 660 patients with nonCOVID-19 viral pneumonia.


In~\cite{pereira2020covid}, the authors also performed a hierarchical analysis for the task of detecting COVID-19 patterns on CXR images. A dataset was built, from other public datasets, containing 1,144 X-ray images, of which only 90 were related to COVID-19 and the remaining belonging to six other classes: five types of pneumonia and one normal (healthy) type. Several techniques were used to extract features from the images, including one based on deep convolutional networks (Inception-V3~\cite{szegedy2016rethinking}). For classification, the authors explored classifiers such as SVM, Random Forest, KNNs, MLPs, and Decision Trees. A F1-Score of 0.89 for the COVID-19 class is reported. In spite of having a strong relation to the present work, we emphasize that a direct comparison is not possible, due to the different nature of the datasets employed on both works.

In~\cite{coronet:20}, the authors propose a Convolutional Neural Network-based model to automate the detection of COVID-19 infection from chest X-ray images, named CoroNet. The proposed model uses the Xception CNN architecture \cite{xception:17}, pre-trained on ImageNet dataset \cite{imagenet_cvpr09}. CoroNet was trained and tested on the prepared dataset from two different publically available image databases (available at \url{https://github.com/ieee8023/covid-chestxray-dataset} and \url{https://www.kaggle.com/paultimothymooney/chest-xray-pneumonia}). The CoroNet model achieved an accuracy of 89.6\%, with precision and recall rate for COVID-19 cases of 93\% and 98.2\% for 4-class cases (COVID vs Pneumonia bacterial vs Pneumonia viral vs Normal) with a 4-fold cross-validation scheme. 
Also, the authors evaluate their model on a second dataset, tough this second dataset apparently contains the same COVID-19 images used during training.






\section{Problem Setting}
\label{sec:setting}

\begin{figure}[!tb]
    \centering
    \begin{subfigure}{.56\columnwidth}
        \centering
        \includegraphics[width=.95\columnwidth]{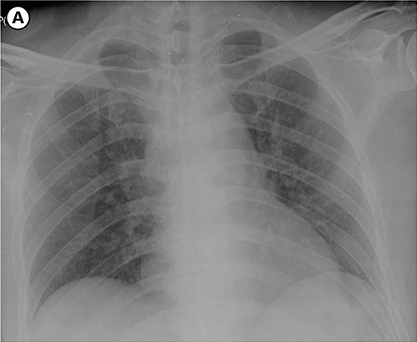}
        \caption{ }
        \label{fig:sub1}
    \end{subfigure}%
    \begin{subfigure}{.42\columnwidth}
        \centering
        \includegraphics[width=.95\columnwidth]{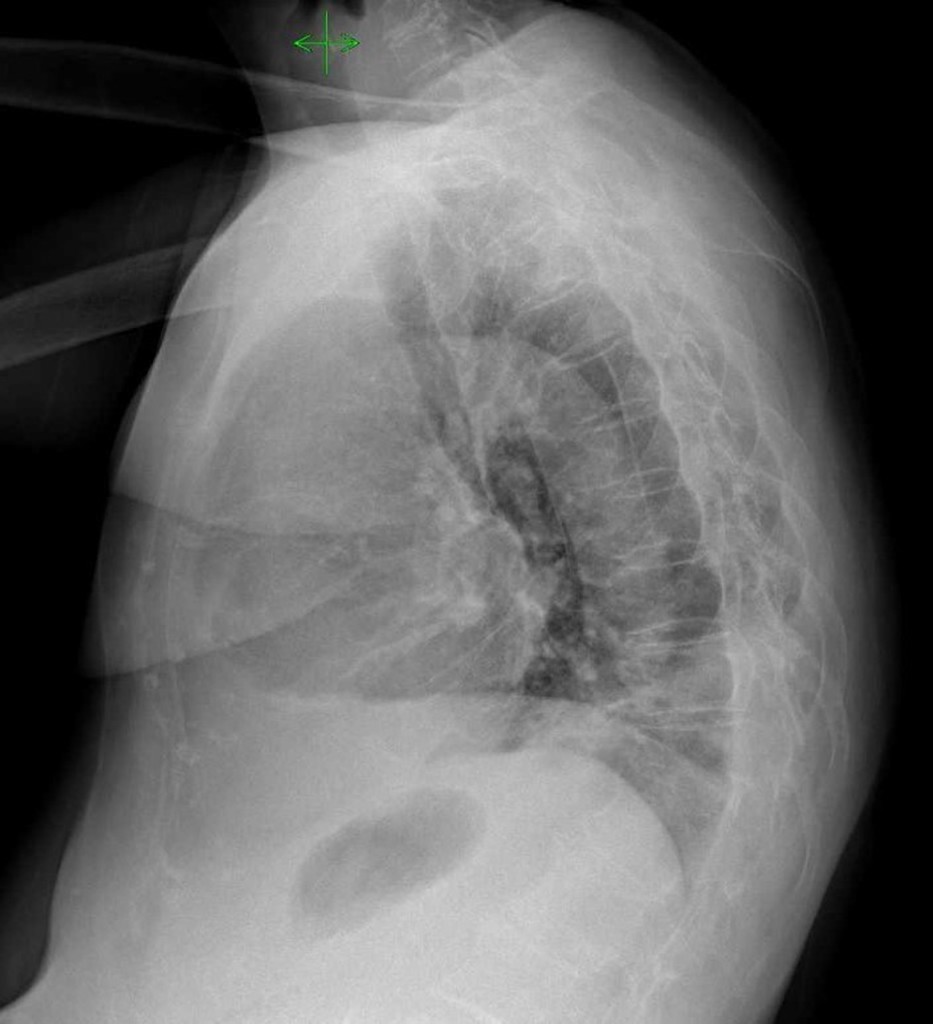}
        \caption{}
        \label{fig:sub2}
    \end{subfigure}
    \caption{Radiograph example of images from COVID-19 image data collection~\cite{cohen2020covid}. (a) X-ray of a 54-year-old male, infected with COVID-19~\cite{cohen2020covid}. (b) X-ray a 70-year-old female, infected with COVID-19~\cite{cohen2020covid}.}
    \label{fig:test}
\end{figure}

The problem addressed by the proposed approach can be defined as follows: Given a chest X-ray, determine if it belongs to a healthy patient, a patient with COVID-19, or a patient with other forms of pneumonia. FIGURE~\ref{fig:test} shows typical chest X-ray samples in COVID$x$ dataset~\cite{wang2020covidnet}. As can be seen, the model should not make assumptions regarding the view in which the X-ray was taken. 

Thus, given an image similar to these ones, a model must output one of the following three possible labels: 
\begin{itemize}
    \item \textit{normal} - for healthy patients.
    \item \textit{COVID-19} - for patients with COVID-19.
    \item \textit{pneumonia} - for patients with non-COVID-19 pneumonia.  
\end{itemize}

Following the rationale in \cite{wang2020covidnet}, choosing these three possible predictions can help clinicians in deciding who should be prioritized for PCR testing for COVID-19 case conﬁrmation. Moreover, it might also help in treatment selection since COVID-19, and non-COVID-19 infections require different treatment plans.

We analyze the problem from two perspectives: 1) the traditional flat classification, in which we disregard the relationship between the classes; and 2) the hierarchical classification approach, in which we assume the classes to be predicted are naturally organized into a taxonomy.

\section{Methodology}
\label{sec:approach}

In this section, we present the methodology for COVID-19 detection by means of a chest X-ray image. 
We detail the main datasets and briefly describe the COVID-Net~\cite{wang2020covidnet}, our baseline method. 
Also, we describe the employed deep learning techniques as well as the learning methodology and evaluation.

\subsection{Datasets}

\subsubsection{RSNA Pneumonia Detection Challenge dataset}
The RSNA Pneumonia Detection Challenge~\cite{RSNAdataset} is a competition that aims to locate lung opacities on chest radiographs. 
Pneumonia is associated with opacity in the lung, and some conditions such as pulmonary edema, bleeding, volume loss, lung cancer can also lead to opacity in lung radiography.
Finding patterns associated with pneumonia is a hard task. 
In that sense, the Radiological Society of North America (RSNA) has promoted the challenge, providing a rich dataset. Although The RSNA challenge is a segmentation challenge, here we are using the dataset for a classification problem. The dataset offers images for two classes: Normal and Pneumonia (non-normal). We are using a total of $16,680$ images of this dataset, of which $8,066$ are from Normal class and $8,614$ from the Pneumonia class.



\subsubsection{COVID-19 image data collection}
The ``COVID-19 Image Data Collection''~\cite{cohen2020covid} is a collection of anonymized COVID-19 images, acquired from websites of medical and scientific associations~\cite{giovagnoni2020facing,sirm2020} and research papers. 
The dataset was created by researchers from the University of Montreal with the help of the international research community to assure that it will be continuously updated. Nowadays, the dataset includes more than 183 X-ray images of patients who were affected by COVID-19 and other diseases, such as MERS, SARS, and ARDS. 
The dataset is public and also includes CT scans images.
According to the authors, the dataset can be used to assess the advancement of COVID-19 in infected individuals, and also allow the identification of patterns related to COVID-19 helping in differentiating it from other types of pneumonia. 
Besides, CXR images can be used as an initial screening for the COVID-19 diagnostic processes. 
So far, most of the images are from male individuals (approx. 60/40\% of males and females, respectively), and the age group that concentrates most cases is from 50 to 80 years old.

\subsubsection{COVIDx dataset}

In \cite{wang2020covidnet}, a new dataset is proposed by merging two other public datasets: ``RSNA Pneumonia Detection Challenge dataset'' and ``COVID-19 Image Data Collection''. 
The new dataset, called COVID$x$, is designed for a classification problem and contemplates three classes: Normal, Pneumonia, and COVID-19. 
Most instances of the Normal and Pneumonia classes come from the ``RSNA Pneumonia Detection Challenge dataset'', and all instances of the COVID-19 class come from the ``COVID-19 Image Data Collection''.
The dataset has a total of $13,800$ images from $13,645$ individuals and is split into two partitions, one for training purposes and one for testing (model evaluation). The distribution of images between the partitions is shown in TABLE~\ref{tab:covidx}, and the source code to reproduce the dataset is publicly available\footnote{https://github.com/lindawangg/COVID-Net}.

\begin{table}[!htb]
    \centering
    \caption{COVID$x$ Images distribution among classes and partitions. The dataset is proposed in~\cite{wang2020covidnet}.}
    \label{tab:covidx}
    \begin{tabular}{@{}ccccc@{}}
    \toprule\toprule
        {\centering Type} &
            {\centering Normal} &
            {\centering Pneumonia} &
            {\centering COVID-19} &
            {\centering Total}\\ \midrule
        Train & 7966 & 5421 & 152  & 13569 \\\midrule
        Test  & 100  & 100  & 31 & 231\\
    
    \bottomrule\bottomrule
    \end{tabular}
\end{table}

\subsection{EfficientNet}

\begin{figure}[!htb]
    \centering
    \includegraphics[width=0.5\linewidth]{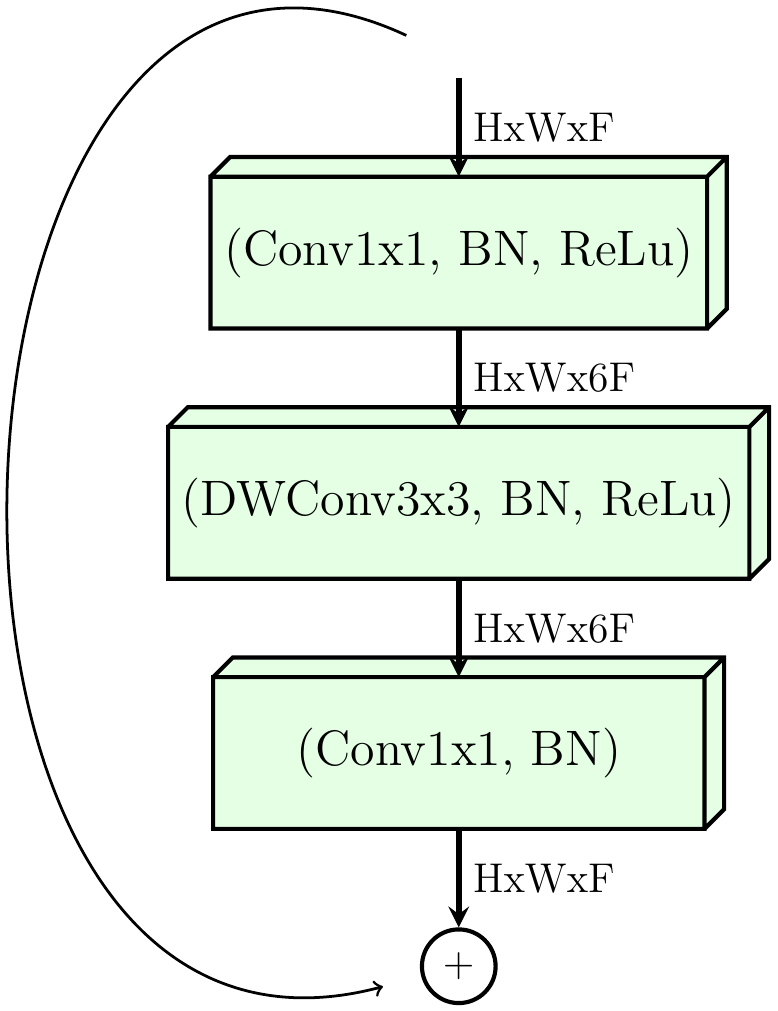}
    \caption{MBConv Block~\cite{sandler2018mobilenetv2}. DWConv stands for depthwise conv,
k3x3/k5x5 defines the kernel size, BN is batch norm, $HxWxF$ means tensor shape (height, width, depth), and ×1/2/3/4 is the multiplier for number of repeated layers. (Figure created by the authors)}
    \label{fig:effblock}
\end{figure}

The EfficientNet \cite{tan2019efficientnet} is in fact a family of models defined on the baseline network described in TABLE~\ref{tab:effNetb0}.
Its main component is known as the Mobile Inverted Bottleneck Conv (MBconv) Block introduced in \cite{sandler2018mobilenetv2} and depicted in FIGURE~\ref{fig:effblock}.  
 
\begin{table}[!ht]
    \centering
    \caption{EfficientNet baseline network : B0 architecture.}
    \label{tab:effNetb0}
    \begin{tabular}{@{}ccccc@{}}
    \toprule\toprule
        \parbox{1cm}{\centering Stage} &
            {\centering Operator} &
            {\centering Resolution} &
            {\centering \#channels} &
            {\centering \#layers}\\ \midrule
        1 & Conv3x3      & 224x224 & 32  & 1 \\\midrule
        2 & MBConv1,k3x3 & 112x112 & 16 & 1\\\midrule
        3 & MBConv6,k3x3 & 112x112 & 24 & 2\\\midrule
        4 & MBConv6,k5x5 & 56x56 & 40 & 2\\\midrule
        5 & MBConv6,k3x3 & 28x28 & 80 & 3\\\midrule
        6 & MBConv6,k5x5 & 14x14 & 112 & 3\\\midrule
        7 & MBConv6,k5x5 & 14x14 & 192 & 4\\\midrule
        8 & MBConv6,k3x3 & 7x7 & 320 & 1\\\midrule
        9 & Conv1x1/Pooling/FC & 7x7 & $1,280$ & 1\\
    \bottomrule\bottomrule
    \end{tabular}
\end{table}

\begin{figure}[!ht]
    \centering
    \includegraphics[width=0.7\linewidth]{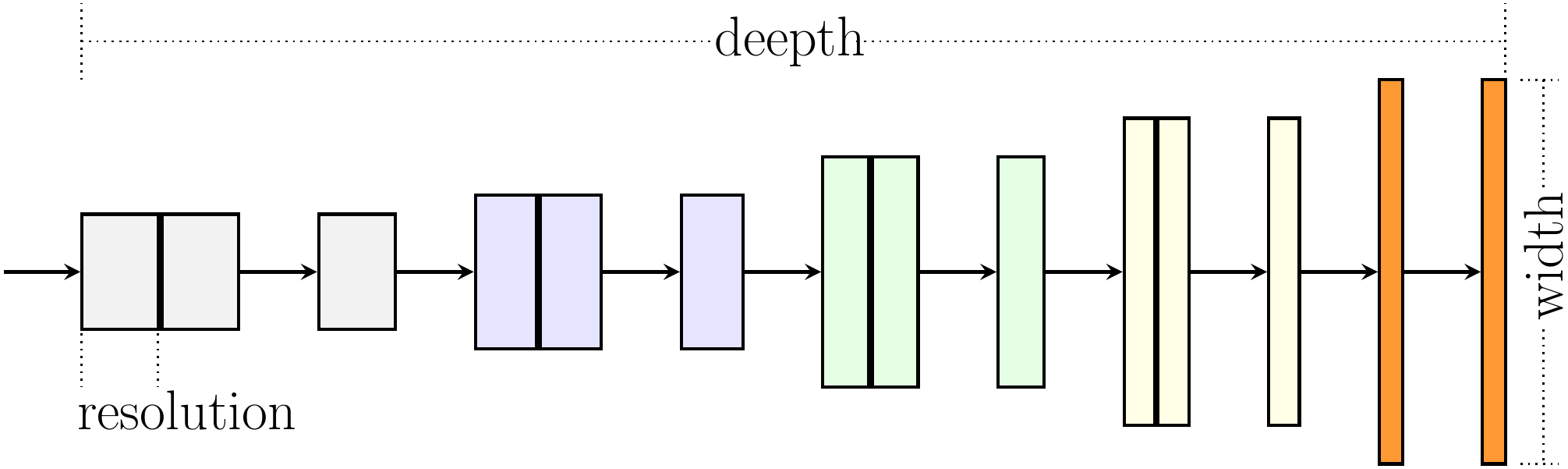}
    \caption{Efficient net compound scaling on three parameters. (Adapted from \cite{tan2019efficientnet})}
    \label{fig:deepernets}
\end{figure}

The rationale behind the EfficientNet family is to start from high quality yet compact baseline model and uniformly scale each of its dimensions systematically with a fixed set of scaling coefficients. Formally, an EfficientNet is defined by three dimensions: (i) depth; (ii) width; and (iii) resolutions as illustrated in FIGURE~\ref{fig:deepernets}.

Starting from the baseline model in TABLE~\ref{tab:effNetb0} each dimension is scaled by the parameter $\phi$ according to
\begin{equation}\label{eq:effnet}
    \begin{split}
        depth      &= \alpha^{\phi }\\  
        width      &= \beta^{\phi }  \\
        resolution &= \gamma^{\phi }  \\
        s.t.       & ~ \alpha  \cdot \beta^{2} \cdot \gamma^{2}\approx 2\\
                   & ~ \alpha\geq  1, \beta\geq 1, \gamma\geq 1\\
    \end{split},
\end{equation}

\noindent where $\alpha$, $\beta$, $\gamma$ are constants obtained by a grid search experiment. 
As stated in \cite{tan2019efficientnet}, Eq.~\ref{eq:effnet} provides a nice balance between performance and computational cost.
The coefficient $\phi$ controls the available resources. 
Eq.~\ref{eq:effnet} determines the increase or decrease of model FLOPS when depth, width or resolution are modified.

Notably, in \cite{tan2019efficientnet}, a model from EfficientNet family was able to beat the powerful GPipe Network~\cite{huang2019gpipe} on the ImageNet dataset~\cite{russakovsky2015imagenet} running with 8.4x fewer parameters and being 6.1x faster.

\subsection{Hierarchical Classification}
\label{subsec:hierarchical}

In classification problems, it is common to have some sort of relationship among classes. Very often, on real problems, the classes (the category of an instance) are organized hierarchically, like a tree structure. According to Silla Jr. and Freitas~\cite{silla2011survey}, one can have three types of classification: flat classification, which ignores the hierarchy of the tree; local classification, in which there is a set of classifiers for each level of the tree (one classifier per node or level); and finally, global classification, in which one single classifier is built with the ability to classify any node in the tree, besides the leaves.

\begin{figure}[!htb]
    \centering
    \includegraphics[width=0.7\linewidth]{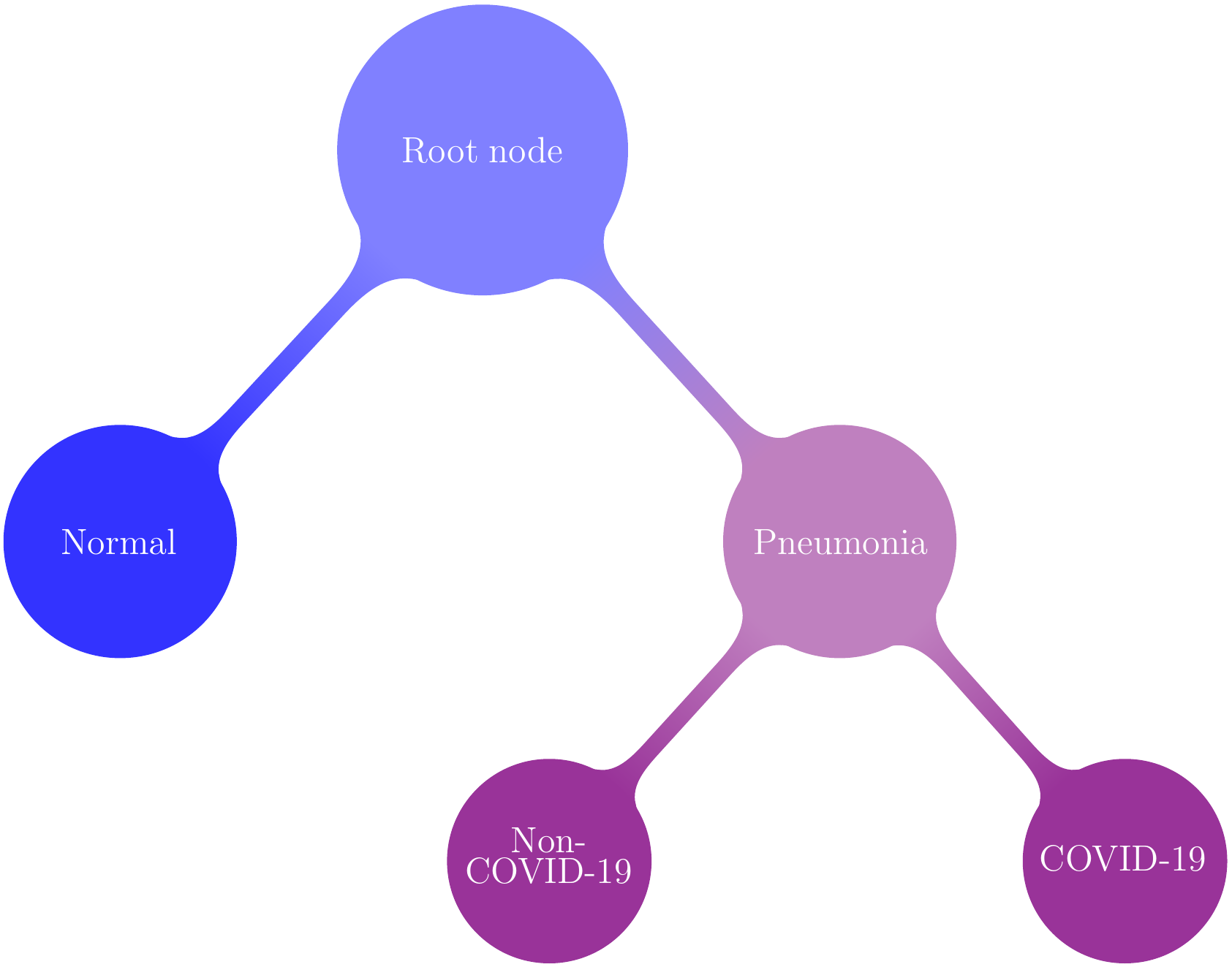}
    \caption{Natural topology of the classes: Normal, Pneumonia, COVID-19. It illustrates the Local-Per-Node hierarchical approach, in which there is a classifier on each parent node. (Figure created by the authors)}
    \label{fig:topology}
\end{figure}

The most popular type of classification in the literature is the flat one. However, here we propose the use of local classification, which we call hierarchical classification. Thus, the target classes are located in the leaves of the tree, and in the intermediate nodes, we have classifiers. In this work, we need two classifiers, one at the root node, dedicated to discriminate between the Normal and Pneumonia classes, and another one in the next level dedicated to discriminate between pneumonia types. 
The problem addressed here can be mapped as the topology depicted in FIGURE~\ref{fig:topology} in which there are two levels of classification. 
To make the class inference for a new instance, first, the instance is presented to the first classifier (in the root node).
If it is predicted as ``Normal'', the inference ends there. If the instance is considered ``Pneumonia'', it is then presented to the second classifier, which will discern whether it is a Pneumonia caused by ``COVID-19'' or ``Not''. 

\subsection{Training}

Deep learning models are complex, and therefore require a large number of instances to avoid overfitting, i.e., when the learned network performs well on the training set but underperform on the test set. Unfortunately, for most problems in real-world situations, data is not abundant. 
In fact, there are few scenarios in which there is an abundance of training data, such as the ImageNet~\cite{russakovsky2015imagenet}, in which there are more than 14 million images of $21,841$ classes/categories.
To overcome this issue, researchers rely on two techniques: data augmentation and transfer learning. 
We also detail here the proposed models, based on EfficientNet.

\subsubsection{Image Pre-processing and Data Augmentation}

Several pre-processing techniques may be used for image cleaning, noise removal, outlier removal, etc. 
The only pre-processing applied in this work is a simple intensity normalization of the image pixels to the range $[0, 1]$.
In this manner, we rely on the filters of the convolutional network itself to perform possible data cleaning. Also, all images are resized according to the architecture resolution parameter (See table~\ref{tab:effNetb0}).

\begin{figure}
    \centering
    \includegraphics[width=0.7\linewidth, trim={3cm 0 3cm 0}, clip]{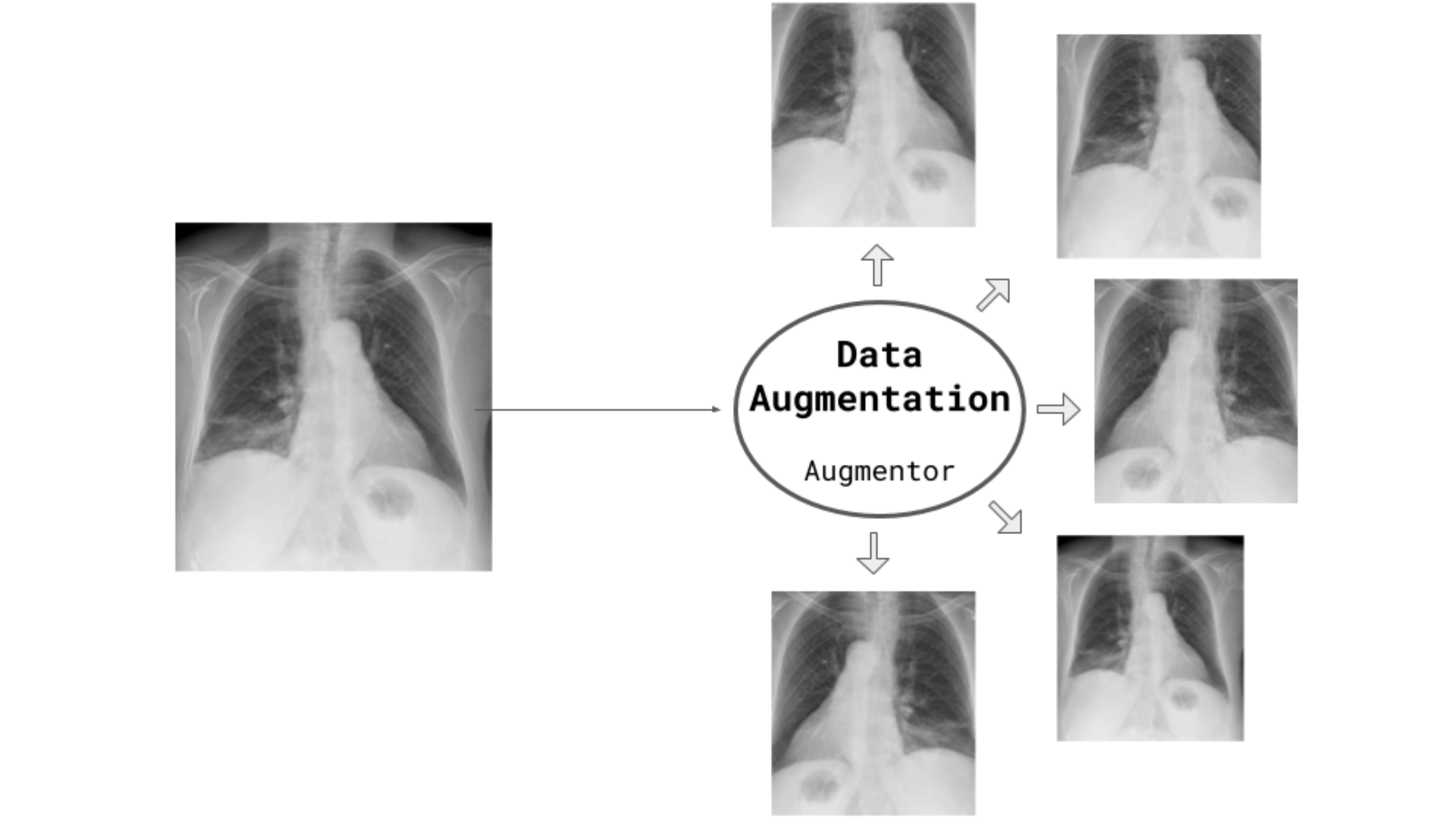}
    \caption{Data augmentation applied using the Augmentor python package. The transformations applied to the images are: rotation (0 to 15 degrees clockwise or anticlockwise), 20\% Zoom or horizontal flipping . All or none changes may be applied/combined according to a probability. (Figure created by the authors)}
    \label{fig:da}
\end{figure}

Data augmentation consists of expanding the training set with transformations of the images in the dataset~\cite{goodfellow2016deep} provided that the semantic information is not lost.  In this work, we applied three transformations to the images: rotation, horizontal flip, and scaling, as such transformations would not hinder, for example, a physician to interpret the radiography. Figure~\ref{fig:da} presents an example of the applied data augmentation.

\subsubsection{Proposed models}
The EfficientNet family has models of high performance and low computational cost. 
Since this research aims to find efficient models capable of being embedded in conventional smartphones, the EfficietNet family is a natural choice.
We explore the EfficientNets by adding more operator blocks atop of it. 
More specifically, we add four new blocks, as detailed in TABLE~\ref{tab:proposed}. 

Since the original EfficientNets were built to work on a different classification problem we add new fully connected layers (FC) responsible for the last steps of the classification process. 
We also use batch normalization (BN), dropout, and swish activation functions for the following reasons.

The \textit{batch normalization} constrains the output of the last layer in a range, forcing zero mean and standard deviation one. That acts as regularization, increasing the stability of the neural network, and accelerating the training~\cite{ioffe2015batch}. 

The \textit{Dropout}~\cite{srivastava2014dropout} is perhaps the most powerful method of regularization. 
The practical effect of dropout operation is to emulate a bagged ensemble of multiple neural networks by inhibiting a few neurons, at random, for each mini-batch during training. The number of inhibited neuronal units is defined by the dropout parameter, which ranges between 0 to 100 percent.

The most popular activation function is the Rectified Linear Unit (ReLU), which can be formally defined as $f(x)=max(0,x)$. 
However, in the added block we have opted for the \textit{swich activation function}~\cite{ramachandran2017searching} defined as: 
\begin{equation}
 f(x) = x \cdot (1+\exp^{-x})^{-1}.
\end{equation}
%
Differently from the ReLU the swish activation produces a smooth curve during the minimization loss process when a gradient descent algorithm is used.
Another advantage of the swish activation regarding the ReLU, it does not zero out small negative values which may still be relevant for capturing patterns underlying the data~\cite{ramachandran2017searching}. 

\begin{table}[!htb]
    \centering
    \caption{Proposed architectures, considering the EfficientNet B0 as base model. (NC = Number of Classes).}
    \label{tab:proposed}
    \begin{tabular}{@{}ccccc@{}}
    \toprule\toprule
        \parbox{1cm}{\centering Stage} &
            {\centering Operator} &
            {\centering Resolution} &
            {\centering \#channels} &
            {\centering \#layers}\\ \midrule
        1-9 & EfficientNet B0 & 224x224 & 32  & 1 \\\midrule
        10 & BN/Dropout & 7x7 & 1280 & 1\\\midrule
        11 & FC/BN/Swich/Dropout & 1 & 512 & 1\\\midrule
        12 & FC/BN/Swich & 1 & 128 & 1\\\midrule
        13 & FC/Softmax & 1 & NC & 1\\
    \bottomrule\bottomrule
    \end{tabular}
\end{table}

\subsubsection{Transfer learning}
Instead of training a model from scratch, one can take advantage of using the weights from a pre-trained network and accelerate or enhance the learning process. 
As discussed in \cite{oquab2014learning}, the initial layers of a model can be seen as feature descriptors for image representation, and the latter ones are related to instance categories. 
Thus, in many applications, several layers can be re-used. The task of transfer learning is then to define how and what layers of a pre-trained model should be used. This technique has proved to be effective in several computer vision tasks, even when transferring weights from completely different domains~\cite{goodfellow2016deep, luz2018deep}. 

The steps for transfer of learning are: 
\begin{enumerate}
    \item Copying the weights from a pre-trained model to a new model;
    \item Modifying the architecture of the new model to adapt it to the new problem, possibly including new layers;
    \item Initialize the new layers;
    \item Define which layers will pass through a new the learning process; and
    \item training (updating the weights according to the loss function) with a suitable optimization algorithm.
\end{enumerate}
 
We apply transfer learning to EfficientNets pre-trained on the ImageNet dataset~\cite{russakovsky2015imagenet}. 
It is clear that the ImageNet domain is much broader than the chest X-rays that will be presented to the models in this work.
Thus, the imported network weights are taken just as an initial solution and are all fine-tuned (i.e., the weights from all layers)  by the optimizer over the new training phase. 
The rationale is that the imported models already have a lot of knowledge about all sorts of objects. 
By permitting all the weights to get fine-tuned we allow the model to specialize to the problem in hands. 
In the training phase, the weights are updated with the Adam Optimizer and a schedule rule decreasing the learning rate by a factor of 10 in the event of stagnation ('patience=2'). 
The learning rate started with $10^{-4}$, and the number of epochs fixed at 10.

\subsection{Model evaluation and metrics}

The final evaluation is carried out with the COVID$x$ dataset, and since the COVID$x$ comprises a combination of two other public datasets, we follow the script\footnote{https://github.com/lindawangg/COVID-Net} provided in~\cite{wang2020covidnet} to load the training and test sets. 
The data is then distributed according to the TABLE~\ref{tab:covidx}.



In this work, three metrics are used to evaluate models: accuracy ($Acc$), COVID-19 sensitivity ($Se_C$), and COVID-19 positive prediction ($+P_C$), i.e.,
\begin{equation}\label{eq:metrics}
    \begin{split}
        Acc &= \frac{TP_N + TP_P + TP_C}{\#samples} \\  & \\
        Se_C &= \frac{TP_C}{TP_C + FN_C}  \\
        & \\
        +P_C &= \frac{TP_C}{TP_C + FP_C}  \\
    \end{split}
\end{equation}
\noindent wherein $TP_N$, $TP_P$, $TP_C$, $FN_C$, and $FP_C$  stand for  
the normal samples correctly classified, 
non-COVID-19 samples correctly classified, 
the COVID-19 samples correctly classified, 
the COVID-19 samples classified as normal or non-COVID-19,  the non-COVID-19 and normal samples classified as COVID-19. The number of multiply-accumulate (MAC) operations are used to measure the computational cost.


\section{Experiments and Discussion}
\label{sec:experiments}

In this section, we present the experimental setup and results for both flat and hierarchical approaches.
The execution environment of the computational experiments was conducted on an Intel(R) Core(TM) i7-5820K CPU @ 3.30GHz, 64Gb Ram, two Titan X with 12Gb, and the TensorFlow/Keras framework for Python.

\subsection{Dataset setup}

Three different training set configurations were analyzed with the COVID$x$ dataset: 
$i)$ (Raw Dataset) -  the raw dataset without any pre-processing; 
$ii)$ (Raw Dataset + Data Augmentation) - the raw dataset with a data augmentation of 1,000 new images on COVID-19 samples and a limitation of 4,000 images for the two remaining classes; and 
$iii)$ (Balanced Dataset) - the dataset with a 1,000 images per class achieved by data augmentation on COVID-19 samples and undersampling the other two classes to $1,000$ samples each one.
Learning with an unbalanced dataset could bias the prediction model towards the classes with more samples, leading to inferior classification models.

In this work, we evaluate two scenarios: flat and hierarchical. 
Regardless of the scenarios, the three training sets remain the same (Raw, Raw + Data Augmentation, and Balanced). 
However, for the hierarchical case, there is an extra process to split the sets into two parts: the first part, the instances of Pneumonia and COVID-19 classes are joined and receive the same label (Pneumonia).
In the second part, the instances related to the Normal class are removed, leaving in the set only instances related to Pneumonia and COVID-19.
Thus, two classifiers are built for the hierarchical case, and each one works with a different set of data (see Section~\ref{subsec:hierarchical} for more details).


\subsection{Experimental settings and results}

\begin{table}[!th]
    \centering
    \caption{Base models footprint details. (Mb = Megabytes)}
    \label{tab:nets}
    \begin{tabular}{@{}cccc@{}}
    \toprule\toprule
        Model           & Input shape & \#Params    & Memory usage (Mb) \\ \midrule
        EfficientNet B0 & 224, 224, 3 & 5,330,564   & 21       \\ \midrule
        EfficientNet B1 & 240, 240, 3 & 7,856,232   & 31       \\ \midrule
        EfficientNet B2 & 260, 260, 3 & 9,177,562   & 36       \\ \midrule
        EfficientNet B3 & 300, 300, 3 & 12,320,528  & 48       \\ \midrule
        EfficientNet B4 & 380, 380, 3 & 19,466,816  & 76       \\ \midrule
        EfficientNet B5 & 456, 456, 3 & 30,562,520  & 118      \\ \midrule
        MobileNet       & 224, 224, 3 & 4,253,864   & 17       \\ \midrule
        MobileNet V2    & 224, 224, 3 & 3,538,984   & 14       \\ \midrule
        ResNet 50       & 224, 224, 3 & 25,636,712  & 99       \\ \midrule
        VGG-16          & 224, 224, 3 & 138,357,544 & 528      \\ \midrule
        VGG-19          & 224, 224, 3 & 143,667,240 & 549      \\
    \bottomrule\bottomrule
    \end{tabular}
\end{table}

We evaluate four families of convolutional neural networks: EfficientNet, MobileNet, VGG and ResNet.
Their features are summarized in TABLE~\ref{tab:nets}.
Among the presented models, we highlight the low footprint of MobileNet and EfficientNet.

Regarding the base models (B0-B5 models of EfficientNet family), the simplest one is the EfficientNet-B0. Thus, we assess the impact of the different training sets and the two forms of classification (flat and hierarchical) from this one. 
The results are shown in TABLE~\ref{tab:efnb0}. 
\begin{table}[!ht]
    \centering
    \caption{EfficientNet B0 results over the three proposed training sets. ($Acc.$ = Accuracy; $Se_C$ = COVID-19 Sensitivity; $+P_C$ = COVID-19 Positive Prediction.)}
    \label{tab:efnb0}
        \begin{tabular}{@{}ccccc@{}}
        \toprule\toprule
            \parbox{1.5cm}{\centering Approach} & Training dataset 
                & \parbox{1.1cm}{\centering $Acc.$} 
                & \parbox{1.1cm}{\centering $Se_C$} 
                & \parbox{1.1cm}{\centering $+P_C$} \\ \midrule
            
                & \parbox{3cm}{\centering Raw Dataset} 
                    & 92.2\% & 67.7\% & 100.0\% \\ \cmidrule{2-5}
               \multirow{1}{*}{Flat} & \parbox{3cm}{\centering Raw Dataset + \\Data Augmentation}         
                    & 93.0\% & 83.8\% & 100.0\% \\ \cmidrule{2-5}
                & \parbox{2cm}{\centering Balanced Dataset} 
                    & 90.0\% & 93.5\% & 100.0\%\\ \midrule
            
                & \parbox{2cm}{\centering Raw Dataset} 
                    & 54.1\% & 83.8\% & 100.0\% \\ \cmidrule{2-5}
               \multirow{1}{*}{Hierarchical} & \parbox{3cm}{\centering Raw Dataset + \\Data Augmentation}         
                    & 90.4\% & 70.9\% & 91.6\% \\ \cmidrule{2-5}
                & \parbox{3cm}{\centering Balanced Dataset} 
                    & 85.7\% & 93.5\% & 82.8\%\\
        \bottomrule\bottomrule
        \end{tabular}
\end{table}

Since there are more pneumonia, and normal x-ray samples than COVID-19, the neural network learning process tends to improve the classification of the majoritarian classes, since they have more weight for the loss calculation.
This may justify the results obtained by balancing the data.
As described in Section~\ref{subsec:hierarchical}, the hierarchical approach is also evaluated here.
First, classes of COVID-19 and common Pneumonia are combined and presented to the first level of classification (Normal vs Pneumonia). At the second level, another model classifies between pneumonia caused by COVID-19 and other causes.

\begin{figure}[!htb]
    \centering
    \includegraphics[width=0.95\linewidth]{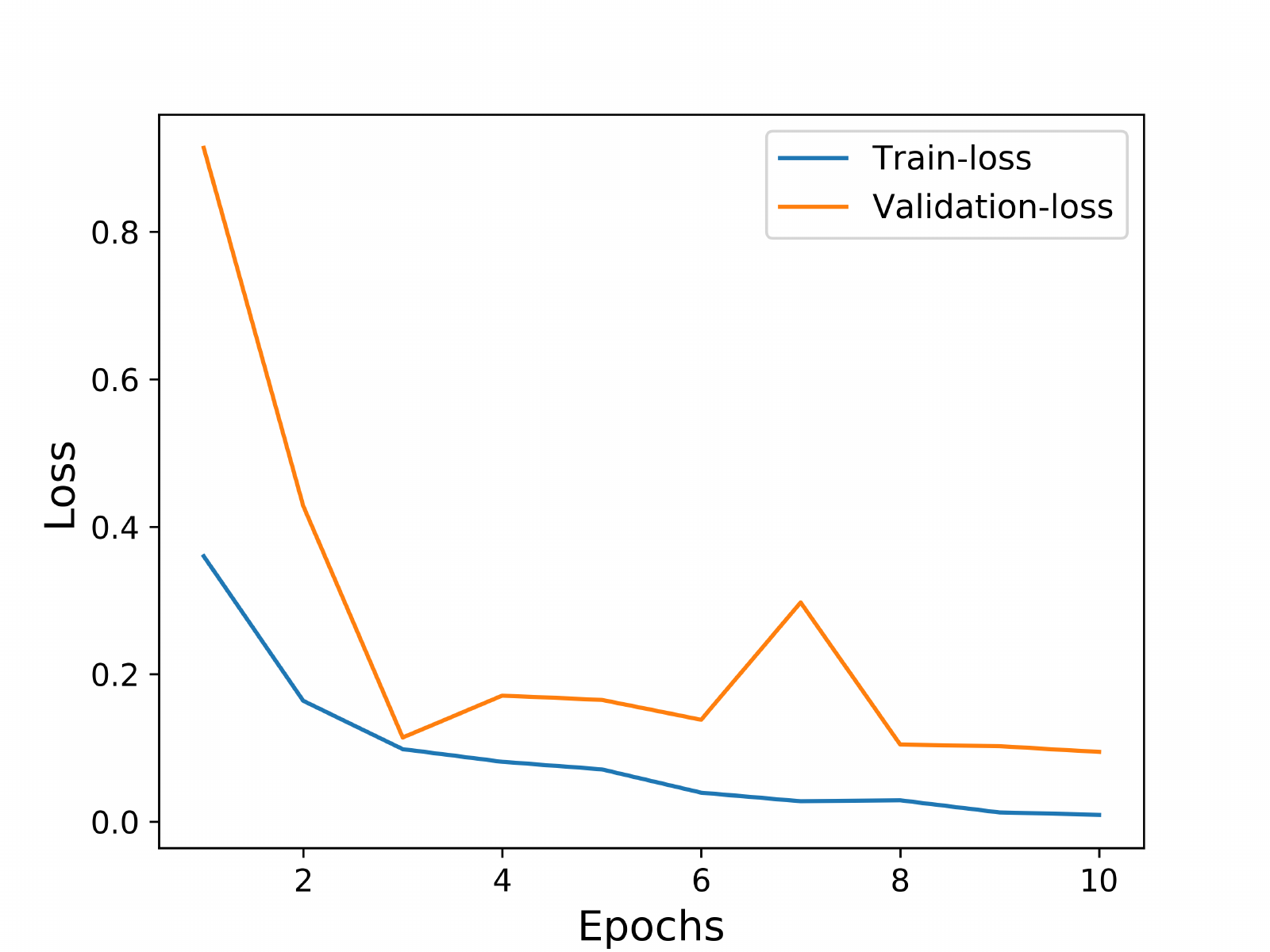}
    \caption{Loss during training-time, EfficientNet-B0 and balanced data. Epochs vs Loss. }
    \label{fig:loss}
\end{figure}

It is possible to see on TABLE~\ref{tab:efnb0} that better results are achieved with the flat approach on balanced data. This scenario is used to evaluate the remaining network as base architectures. The training loss for this scenario is presented in FIGURE~\ref{fig:loss}.

The results of all evaluated architectures are summarized in TABLE~\ref{tab:nets-results}. We stress that we adapted all architectures by placing the same four blocks on top.
It can be seen that all the networks have comparable performances in terms of accuracy. However, the more complex the model is, the worse is the performance for the minority class, the COVID-19 class. 

The cost of a model is related to the number of parameters.
The higher the number of parameters, the higher the amount of data the model needs to adjust them. Thus, we hypothesized that the lack of a bigger dataset may explain the difficulties faced by the more complex models. 


\begin{table}[!htb]
    \centering
    \caption{Results on different network architectures as base model. Best scenario for COVID-19: all experiments with a balanced training set and flat classification. ($Acc.$ = Accuracy; $Se_C$ = COVID-19 Sensitivity; $+P_C$ = COVID-19 Positive Prediction.)}
    \label{tab:nets-results}
    \begin{tabular}{@{}cccc@{}}
    \toprule\toprule
        \parbox{2.5cm}{\centering Base Model  }
            & \parbox{1.5cm}{\centering $Acc.$}
            & \parbox{1.5cm}{\centering $Se_C$}
            & \parbox{1.5cm}{\centering $+P_C$} \\ \midrule
        EfficientNet B0 & 90.0\%   & 93.5\%        & 100.0\%  \\ \midrule
        EfficientNet B1 & 91.8\%   & 87.1\%        & 100.0\%  \\ \midrule
        EfficientNet B2 & 90.0\%   & 77.4\%        & 100.0\%  \\ \midrule
        EfficientNet B3 & 93.9\%   & 96.8\%        & 100.0\%  \\ \midrule
        EfficientNet B4 & 93.0\%   & 90.3\%        & 93.3\%   \\ \midrule
        EfficientNet B5 & 92.2\%   & 93.5\%        & 90.6\%   \\ \midrule
        MobileNet       & 90.4\%   & 83.8\%        & 100.0\%  \\ \midrule
        MobileNet V2    & 90.0\%   & 87.1\%        & 96.4\%   \\ \midrule
        ResNet-50       & 83.5\%   & 70.9\%        & 81.4\%   \\ \midrule
        VGG-16          & 77.0\%   & 67.7\%        & 63.64\%  \\ \midrule
        VGG-19          & 75.3\%   & 77.4\%        & 50.0\%   \\ \midrule
    \bottomrule\bottomrule
    \end{tabular}
\end{table}

TABLE~\ref{tab:sota} presents a comparison of the proposed approach and the one proposed by Wang \textit{et al.}~\cite{wang2020covidnet} (COVID-net) under the same evaluation protocol.
Even though the accuracy is comparable, the proposed approach presents an improvement on positive prediction without losing sensitivity.
Besides, a significant reduction both in terms of memory (our model is >15 times smaller) and latency is observed. It is worth highlighting that Wang \textit{et al.}~\cite{wang2020covidnet} apply data augmentation to the dataset but it is not clear in their manuscript how many new images are created.

\begin{table}[!htb]
    \small
    \centering
    \caption{Comparison of the proposed approach against SOTA. ($Acc.$ = Accuracy; $Se_C$ = COVID-19 Sensitivity; $+P_C$ = COVID-19 Positive Prediction.)}
    \label{tab:sota}
    \begin{tabular}{@{}ccccccc@{}}
    \toprule\toprule
        Method 
            & $Acc.$ & $Se_C$ & $+P_C$ & \parbox{1.4cm}{\centering \#Params (millions)} & \parbox{1.4cm}{\centering MACs (millions)} & \parbox{1.3cm}{\centering Memory required}
    \\ \midrule            
        \parbox{2.2cm}{Approach Flat\\ EfficientNet B3} 
            & 93.9\% & \textbf{96.8\%} & \textbf{100.0\%} & \textbf{11.6} & \textbf{11.5 } & \textbf{134Mb} \\ \midrule
        \parbox{2.2cm}{Approach \\ Hierarchical\\ EfficientNet B3} 
            & 93.5\% & 80.6\% & \textbf{100.0\%} & 23.2 & 23 & 268Mb  \\ \midrule
        \parbox{2.3cm}{COVID-net \cite{wang2020covidnet}} & \textbf{94.3\%} & \textbf{96.8\%} & 90.9\% & 126.6 & 3500 & 2.1Gb\\
    \bottomrule\bottomrule
    \end{tabular}
\end{table}

The COVID-Net~\cite{wang2020covidnet} is a very complex network, which demands a memory of 2.1GB (for the smaller model) and performs over 3.5 billion MAC operations implying three main drawbacks: computation-cost, time-consumption, and infrastructure costs. A 3.59 billion MAC operations model takes much more time and computations than a 11.5 million MAC model - in the order of almost 300 times -, and the same GPU necessary to run one COVID-Net model can run more than 15 models of the proposed approach (based on the EfficienteNet B3 flat approach) keeping a comparable (or even better) figures.
The improvements, in terms of efficiency,  are even greater using the EfficientNet B0 - with a small trade-off in terms of the sensitivity metric.
The complexity can hinder the use of the model in the future, for instance, on mobile phones or common desktop computers (without GPU).


\subsection{Discussion}



\begin{figure}
    \centering
    \begin{tabular}{cc}
        \includegraphics[width=.45\linewidth,page=7]{tex/figs/heatmap.pdf}
            & \includegraphics[width=.45\linewidth,page=8]{tex/figs/heatmap.pdf} \\
        
        \includegraphics[width=0.45\linewidth, page=1]{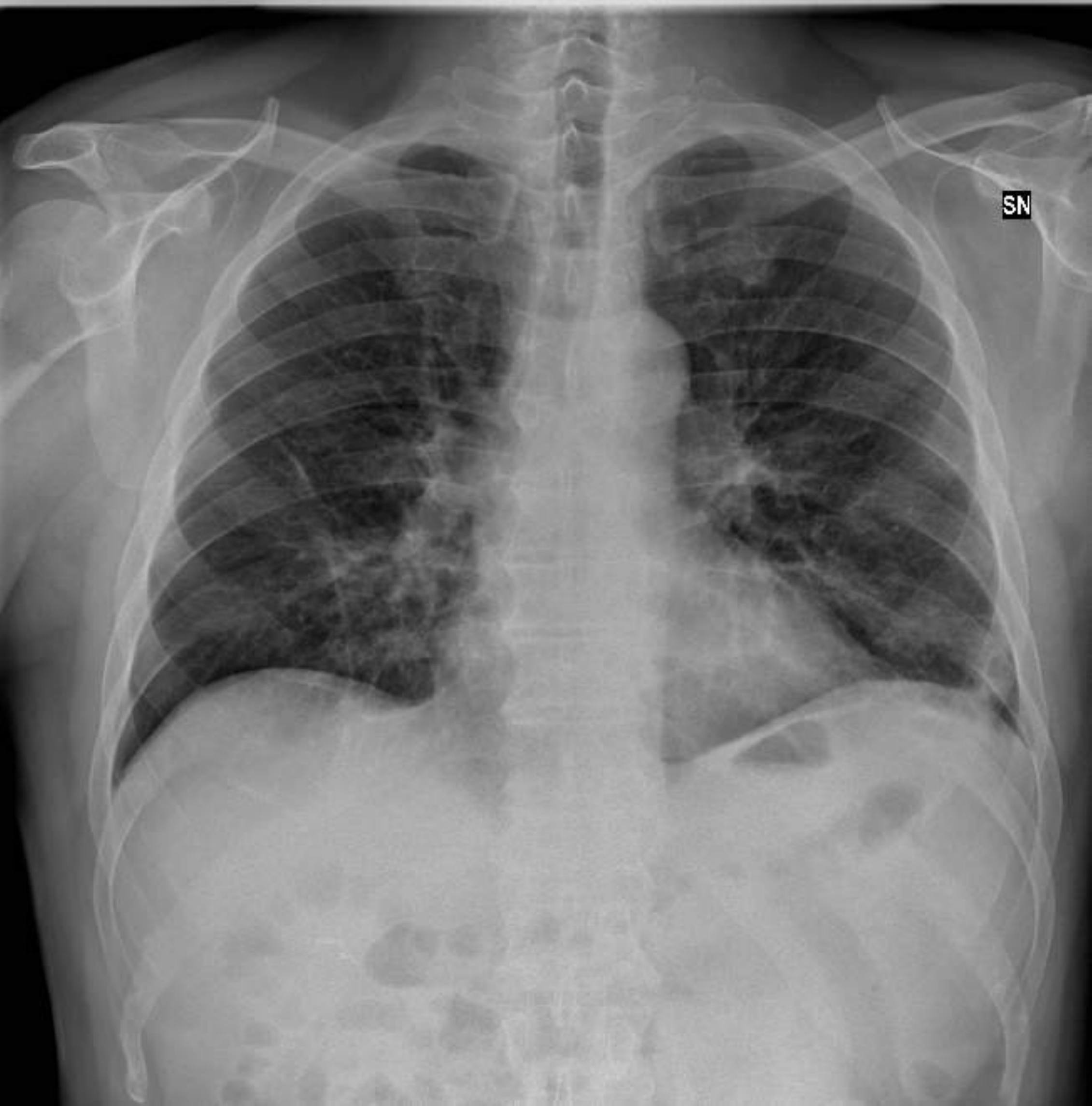}
            & \includegraphics[width=0.45\linewidth, page=2]{tex/figs/activations.pdf} \\
        
    \end{tabular}
    \caption[]{Original images and their activation maps according to the proposed approach. First row presents a patient with COVID-19 (a corrected classified image and its respective activation maps generated by our model), the second, from a healthy chest x-ray sample (the model failed to find opaque points in the image and the activation map highlights non-opaque regions).}
    \label{fig:heatmap}
\end{figure}

In FIGURE~\ref{fig:heatmap}, we present two X-ray images of COVID-19 infected individuals. Those images are from the test set and therefore, were not seen by the model during training. According to studies~\cite{ng2020imaging}, the presence of the COVID-19 infection can be observed through some opacity (white spots) on chest radiography imaging. In the first row of FIGURE~\ref{fig:heatmap}, one can see the corrected classified image and its respective activation maps generated by our model. The activation map corresponds to opaque points in the image, which may correspond to the presence of the disease. For the second row images, it is observed that the model failed to find opaque points in the image and the activation map highlights non-opaque regions.



\begin{figure}[!htb]
    \centering
    \begin{tabular}{@{}cc@{}}
        \includegraphics[
            width=.47\linewidth,
            trim={1cm 0 1cm 0},
            clip,
            page=1]{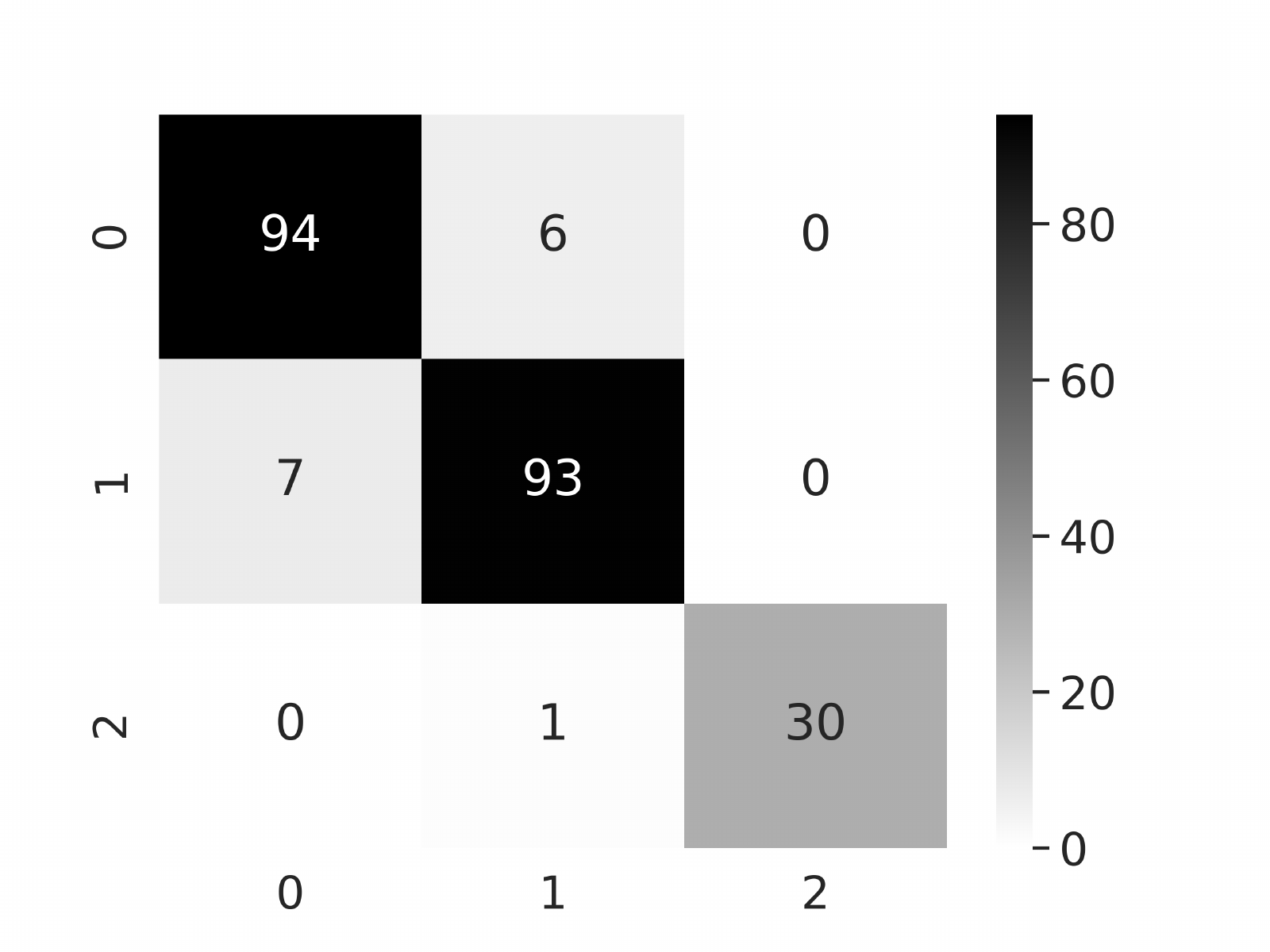} &
        \includegraphics[
            width=.47\linewidth,
            trim={1cm 0 1cm 0},
            clip,
            page=2]{tex/figs/cm.pdf}  \\
    \end{tabular}
    \caption{Confusion matrix of flat (left) and hierarchical (right) approaches respectively with balanced training set. Class zero is the normal images, 1, pneumonia non-COVID-19, and, 2, COVID-19.}
    \label{fig:cm}
\end{figure}

In FIGURE~\ref{fig:cm}, the confusion matrices of flat and hierarchical approaches are presented.  
It is possible to observe that the hierarchical model classifies the normal class better, though it also shown a noticeable reduction in terms of sensitivity and positive prediction for the COVID-19 class.
One hypothesis is that both Pneumonia and COVID-19 classes are similar (both kinds of pneumonia) and share key features.
Thus, the lack of normal images on second classification level reduces the diversity of the training set, interfering with model training. 
Besides, the computational cost is twice as higher than flat classification since two models are required. However, we believe that the hierarchical approach has a key aspect: it suffers less from bias in the dataset/protocol. In~\cite{maguolo2020critic}, a critical evaluation of the test protocols and databases for methods aiming at classifying COVID-19 in X-ray images is presented. According to Maguolo and Nanni~\cite{maguolo2020critic}, the considered datasets are mostly composed of images from different distributions, different databases, and this may favor the deep learning models to learn patterns related to the image acquisition process, instead of focusing only on disease patterns.

\begin{figure}[!htb]
    \centering
    \begin{tabular}{@{}cc@{}}
        \includegraphics[
            width=.47\linewidth,
            trim={1cm 0 1cm 0},
            clip,page=1]{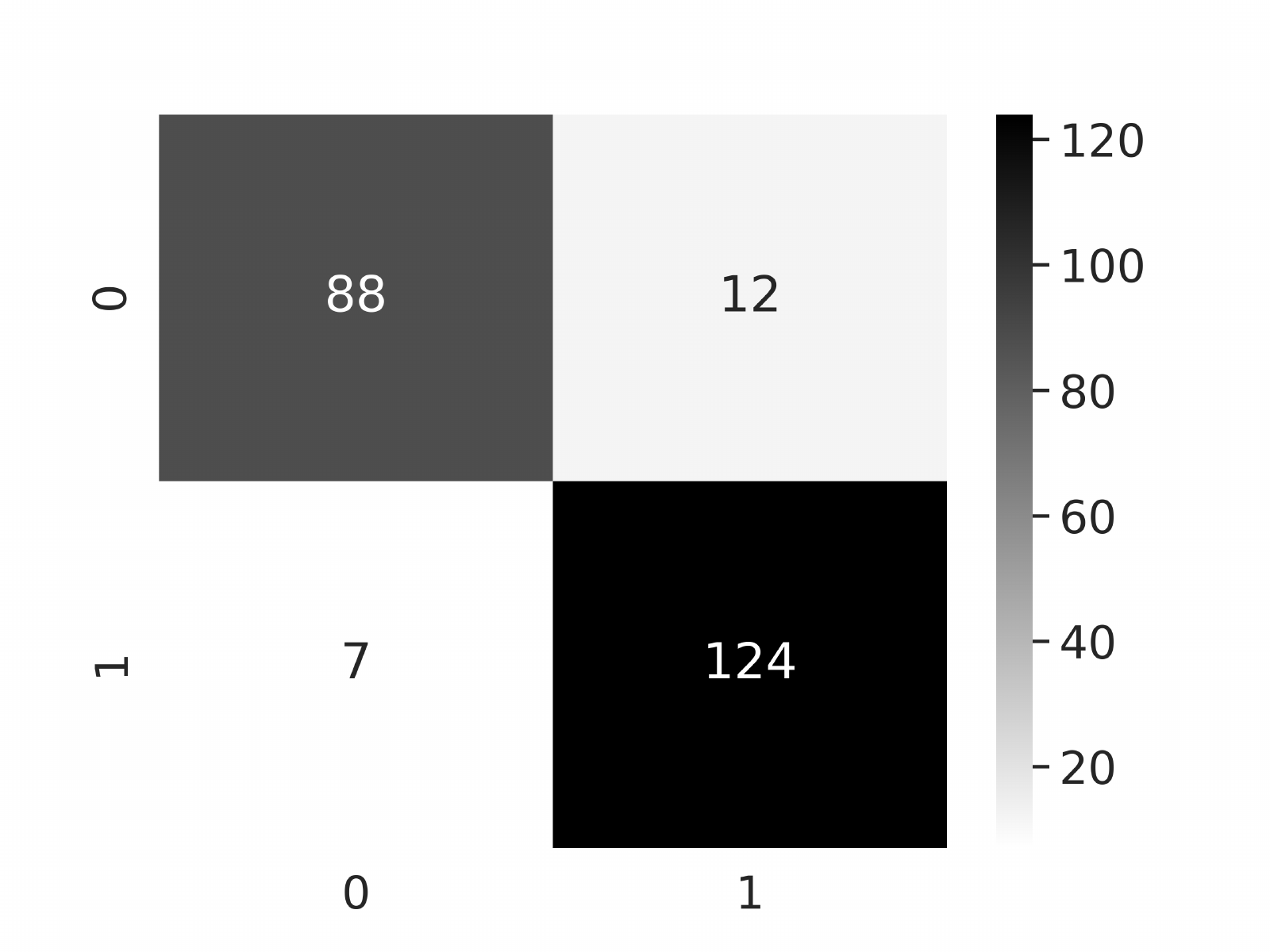} &
        \includegraphics[
            width=.47\linewidth,
            trim={1cm 0 1cm 0},
            clip,page=2]{tex/figs/cm-hierarchical.pdf}  \\
    \end{tabular}
    \caption{Confusion matrix hierarchical approach respectively with balanced training set. On first hierarchical stage (a), class zero is for the normal images and class 1 all Pneumonias (non-covid and COVID-19 pneumonias). On second stage (b) class 0  non-COVID-19 pneumonia, and, 2, COVID-19.}
    \label{fig:cm_hierarch}
\end{figure}

In the first stage of the hierarchical classification, images related to COVID-19 and non-covid pneumonia are given the same classification label. Thus images from different datasets are combined which forces the method to disregard patterns related to the acquisition process or sensors at the first classification stage. An example of the hierarchical model application can be seen in Figure~\ref{fig:cm_hierarch}. It can be seen from the confusion matrix of the first stage (Figure~\ref{fig:cm_hierarch} (a)), that the model is able to classify most instances correctly and for that, we believe it has focused on the patterns that may help discriminate among different types of pneumonia.

\section{Findings and Future Direction}
\label{sec:future}

We summarize our findings as follows.

\begin{itemize}
    \item An efficient and low computational approach was proposed to detect\linebreak COVID-19 patients from chest X-ray images.
    Even with only a few images of the COVID-19 class, insightful results with a sensitivity of 90\% and a positive prediction of 100\% were obtained, with the evaluation protocol proposed in~\cite{wang2020covidnet}.

    \item Regarding the hierarchical analysis, we conclude that there are significant gains that justify the use of the present task. We believe that it suffers less from the bias present in the evaluation protocols, already discussed in~\cite{maguolo2020critic}.
    
    \item The proposed network blocks, put on top of the base models, showed to be very effective for the CRX detection problem, in particular, CRX related to COVID-19.
    
    \item The evaluation protocol proposed in~\cite{wang2020covidnet} is based on the public dataset ``COVID-19 Image Data Collection''~\cite{cohen2020covid}, which is being expanded by the scientific community. 
    With more images from the COVID-19 class, it will be possible to improve the training. However, the test partition tends to become more challenging. 
    For sake of reproducibility and future comparisons of results, our code is available at~\url{https://github.com/ufopcsilab/EfficientNet-C19}.
        

    \item The Internet of Medical Things (IOMT)~\cite{joyia2017internet} is now a hot topic on industry.
    However, the internet can be a major limitation for medical equipment, especially in poor countries. 
    Our proposal is to move towards a model that can be fully embedded in conventional smartphones (edge computing), eliminating the use of the internet or cloud services. 
    In that sense, the model achieved in this work requires only 55Mb of memory and has a viable inference time for a conventional cell phone processor.
\end{itemize}

\section{Conclusion}
\label{sec:conclusion}

In this paper, we exploit an efficient convolutional network architecture for detecting any abnormality caused by COVID-19 through chest radiography images.
Experiments were conducted to evaluate the neural network performance on the COVIDx dataset, using two approaches: flat classification and hierarchical classification. 
Although the datasets are still incipient and, therefore, limited in the number of COVID-19 related images, effective training off the deep neural networks has been made possible with the application of transfer learning and data augmentation techniques.

Concerning evaluation, the proposed approach brought improvements compared to baseline work, with an accuracy of 93.9\%, COVID-19 Sensitivity of 96.8\% and Positivity Prediction of 100\% with a computational efficiency more than 30 times higher.

We believe that the current proposal is a promising candidate for embedding in medical equipment or even physicians' mobile phones. However, larger and more heterogeneous databases are still needed to validate the methods before claiming that deep learning can assist physicians in the task of detecting COVID-19 in X-ray images. 

\section*{Conflict of interest statement}

The authors declare that they have no known competing financial interests or personal relationships that could have appeared to influence the work reported in this paper.


\bibliography{elsarticle-template}

\end{document}